# Detection of Insider Threats using Artificial Intelligence and Visualisation

Vasileios Koutsouvelis*, Stavros Shiaeles†, Bogdan Ghita‡, Gueltoum Bendiab‡
*Open University of Cyprus, 33 Yiannou Kranidioti Ave., Latsia, Nicosia, Cyprus
vasileios.koutsouvelis@st.ouc.ac.cy
†Network and Security Research Group (NSRG), School of Computing,
University of Portsmouth, Winston Churchill Avenue, Portsmouth, PO1 2UP, Hampshire, UK
sshiaeles@ieee.org
‡ Centre for Security, Communications and Networks Research (CSCAN),
School of Computing and Mathematics, Plymouth University, Drake Circus, Plymouth PL4 8AA, UK
bogdan.ghita@plymouth.ac.uk, gueltoum.bendiab@plymouth.ac.uk

*Abstract*—Insider threats are one of the most damaging risk factors for the IT systems and infrastructure of a company or an organization; identification of insider threats has prompted the interest of the world academic research community, with several solutions having been proposed to alleviate their potential impact. For the implementation of the experimental stage described in this study, the Convolutional Neural Network (from now on CNN) algorithm was used and implemented via the Google TensorFlow program, which was trained to identify potential threats from images produced by the available dataset. From the examination of the images that were produced and with the help of Machine Learning, the question whether the activity of each user is classified as "malicious" or not for the Information System was answered.

## I. INTRODUCTION

Computers nowadays are used in every human activity and all kinds of operations, from residential to commercial and from basic service provisioning to research. Beyond their benefit, the risks and cases of deliberate or accidental destruction, tampering or unauthorised use of data and computer resources, in general, are increasing. The consequences of possible tampering, leakage, or misuse of data can lead not only to significant damage and costs but also risks for the protection of the human rights of individuals [1]. In the current security monitoring landscape, artificial neural networks play an extremely important role in the prevention and handling of internal threats [2] and alleviating the risks associated with information systems infrastructures. While a typical infrastructure would have a level of protection against an external attack/threat, an internal threat refers to a person who may have privileged access to classified, sensitive, or proprietary data and uses this advantage to remove information from an organization and transfer it to unauthorised external users. Such attackers may include the users of the company, who can bypass the control procedures for access to classified data, and the users who gain access to user accounts with more rights in relation to these, which they already have. The purpose of this survey was to answer the question of whether Artificial Intelligence can be successfully used to detect malicious activity. The answer to this question went through three (3) steps: (a) collecting, processing, and classifying the data of the users tested; (b) visualizing the extracted data; (c) categorize the behaviour as malicious or normal.

## II. RELATED WORK

The detection of malicious activity with the help of Artificial Intelligence has been a matter of concern to scholars, who have occasionally dealt with this issue, using different approaches. For instance, X. Ren, L. Wang, [3] presented a hybrid insider threat detection system, consisting of data processing, entity portrait, rule matching as well as iterative attention. Based on the results of the experiments, the authors claim that the proposed system provides a higher detection rate and better timeliness since it incorporates multiple complementary detection techniques and components. In [4], Sajjanhar, Atul et al proposed an image-based feature representation of the daily resource usage pattern of users in the organization. authors reported an overall accuracy of 99% when compared with other techniques. In another recent study [5], Kim, Junhong et al introduced an insider threat detection framework based on user behaviour modelling and anomaly detection algorithms and they support those experimental results indicate that the proposed framework can work well for imbalanced datasets in which there are only a few insider threats and where no domain experts' knowledge is provided. In the same context, Hu, Teng et al [6] proposed a continuous identity authentication method based on mouse dynamic behaviour and deep learning to solve the insider threat attack detection problem and they claim that the experimental results showed that the proposed method could identify the user's identities in seconds and has a lower false accept rate (FAR) and false reject rate (FRR).

Tuor, Kaplan and Hutchinson [7] referred to a system of profound knowledge for filtering log files' data and analysing them. According to the writers, an internal threat behaviour varies widely, so the researcher does not attempt to formulate the pattern of behaviour which is a threat. Instead, new variants of Neural Networks (DNN) and Recurring Neural Networks (RNNs) are trained to recognize the activity that is typical for each user on a network. At the same time, these Neural Networks assess whether the behaviour of the user is normal or suspicious. The authors note that detecting malicious cases is particularly difficult because attackers often try to imitate the typical behaviour of a normal user. In another study, Sanzgiri and Dasgupta [8] presented the techniques that have been developed to detect internal threats referring to the researchers and their techniques. In particular, the objective of this paper is to present a categorization of the techniques used by researchers (Hu, Giordano, Kandias, Maybury, Greitzer, Eldardiry) to deal with insider threats. Finally, the researchers remarked that one of the main reasons why it is still difficult to detect attacks by internal users is the lack of sufficient real available data in order to build and test models and mechanisms for detecting internal threats.

Breier and Branisova [9] proposed a threat detection method based on data mining techniques for analysing system log files (log analysis). Their approach was based on Apache Hadoop, which allows the processing of large volumes of data in a parallel way. The method detects new types of violations without further human intervention, while the overall error reaches a value below 10%. Legg, Buckley, Goldsmith and Creese (2015) [10] proposed Corporate Insider Threat Detection (CITD), a corporate threat detection system which incorporates technical and behavioural activities for the evaluation of threats caused by individuals. In particular, the system recognised the users and the role-based profiles and measured how they deviate from their observed behaviour in order to estimate the potential threat that a set of user activities can cause. Some other studies have used approaches [11] based on graphs to find malicious cases in structural data models that represent an internal threat activity that is looking for activities that display similarities to normal data transactions but are structurally different from them. Others [12] have suggested approaches that combine Structural Anomaly Detection - SA.

Despite the work to date, the challenge to holistically observe and analyse user and application behaviour remains a current one, due to its volume and complexity. The benefits of AI-based solution are obvious when faced with the large amounts of data collected, but interpreting the data and results demands further research.

### III. Proposed approach

As identified in the previous section, existing research demonstrated the efficiency of machine learning approaches, but also exposed their limitations in segregating the complexity of user behaviour into normal and malicious. To account for this complexity, we apply the CNN algorithm on user interaction data, as reflected through the log files collected from individual users. Unlike other studies focusing on domain knowledge to detect malicious behaviours through the use of specific structural anomaly detection [13], our proposed approach focuses exclusively on the behaviour of each user of the Information Technology System. In addition to the standard approach of log data collection [10, 14], the method also considers the organizational role for each user when establishing the behaviour profile, which improved significantly the accuracy of the method. For example, the "malicious" activity of a user, who holds an IT Admin position in the organization, may justify - to a certain extent - this activity.

In the context of internal threats, the method aims to discriminate between legitimate and malicious behaviour by investigating the differences in the associated visualisation maps. For each user, the approach creates an image that depicted his/her activity and behaviour, as emerged from their interaction with various information systems. While the resulting images may appear visually different, they were processed through a machine learning algorithm in order to automatically recognize which subset of the users appear to exhibit malicious behaviour (and therefore posing a threat for the respective information systems) and which are legitimate/benign ones.

Two stages are critical for the success of the method: the input data and the processing method. Given the proposed approach is aiming to provide a holistic view of the user interactions, the most appropriate method was considered to be converting log events into a visual representation. A full overview of the process is provided in the following section, including the attempts to highlight the most relevant features through the chosen visualisation. The second critical stage, information processing, was biased by the choice of input data. Research undertaken in recent years demonstrated that Convolutional Neural Networks are indeed extremely efficient at image analysis, as shown by [15]. However, they have also proved their efficiency in the security area. Given their ability to deal with complex relationship, CNNs have been applied from basic security problems, as in [16] which introduced a CNN-based generic detection engine, to [17] for analysis of encrypted content.

The approach presented in this paper follows from the malware classification method from Wang et al. [18] of converting the data analysis task into an image recognition one. Unlike [18] and [19], which look at either malware or network activity to detect attacks, this paper aims to extend the analysis into the logging footprint to detect malicious vs normal behaviour. As highlighted in previous studies, the challenge is to ensure that sufficient input data is available for the method to be successful, and the transformation applied makes it compatible with the image analysis task.

*A. Methodology*

The input data was the Insider Threat Test Dataset from CERT, which is part of the Institute of Software Engineering (SEI). The file included log files, CSV type, which recorded an activity covering eighteen (18) months, collected between 01.01.2010 and 31.05.2011. Through these files and after analysis and processing, it was attempted to present an image of the Information System and to analyse the behaviour of users identified as malicious. For each user, the inputs included login records, files/documents used or opened, emails sent and received, web browsing history, devices used, and user role within the organization.

The steps we followed to complete the process and draw our conclusions were 1) data sharing and creation of files based on the data of the user under consideration, 2) importing the data files we created in the "D3.js" library, selecting an appropriate image creation plan, examining the application library's patterns and creating images of the user in question that included his/her activity during each day, 3) creating images, 4) implementing and training the CNN algorithm in Tensorflow program and examining user behaviour, which we have described as "normal" or "malicious"; and 5) drawing conclusions.

*1) First Stage:* In the first stage of pre-processing, the raw input files were parsed to classify the log files from fifteen users according to the user role (salesman, IT admin, electrical engineer, mechanical engineer, administrator, production line worker, computer scientist, and software quality engineer). First, log files were parsed for each of the 15 users to create separate profiles consisting of three files (file.csv, email.csv, http.csv), which included the complete activity. In the second step, the profile files were compared against a number of rules that defined threats and malicious behaviour. In the website category, we chose terms that were associated with social networks, work search sites, malware, and file sharing. The parsing scripts also searched for attached files, which were divided into three size intervals, defining small [50K B-100K B] medium [100KB-200KB] and large files (over 200KB).

In parallel, we also parsed the profile files for terms associated with malware, such as Keylogger, files that may have been leaked to the Internet, and files that may have beendistributed over the Internet. The result of the pre-processing was a set of three files per user, which had the following structure: date, user (username), host (from which the action was carried out), keyword (defining the type of threat), threat index (each specific threat was assigned a numerical category)which files contained the respective threats. The activity was classified using a separate record field which defined a specifictype of [illegal] activity. At a subsequent stage, the field was transcoded to a unique colour for visualization of the images associated with the user activity map. The collected content for the users was then aggregated into weekly and monthly activity for long-term analysis.

*2) Second Stage:* In the second stage of the experiment, the numerical input was processed through the Java D3.js library to visualize and produce images that depict the activity of each user. Concerning D3.js is an open-source JavaScript library, used for document and data handling, which is based on web templates. It is used through upgraded web browsers, combining powerful data visualization methods and elements. Large datasets can be easily linked to SVG objects usingsimple D3.js functions to create rich text and graphics diagrams. With a minimal resource burden on a computer system, D3.js is extremely fast, supporting large data sets and dynamic interactions through static or moving images.

As mentioned above, the illegal activity was colour-coded for visual processing as follows: blue for social and professional interaction, such as job search and social networking sites, red for sharing site file-sharing websites, pink for email threats, green for file threats, yellow for 50-100 KB email attachments, orange for 100-200 KB email attachments, and brown for $> 200 KB$ email attachments. Using this coding, the user data was converted to an hourly resolution heatmap, summarising week-long and month-long data; this allowed a harmonised view of the data for the two timeframes. This process aimed to convert the numerical data into a visual representation and characterization of the activity to determine whether itis malicious or normal.

The result is a two-dimension heatmap, colour-coded as described above, with the days of the month on the vertical and the hours of the day horizontally. The dimensions of the design area were determined according to the size of the grid that was set. The result was the creation of a space consisting of hourly activity squares, each of which was coloured with the dominant activity identified during that respective timeslot.

Following the generation of the activity-based heatmap, the full dataset included a total of 1199 images; these were manually labelled as normal or malicious. The selection was based on specific criteria, indicative of the density of the activity at specific time intervals, its colours, and the time of day. Following manual analysis, 769 images were labelled as containing malicious activity and 430 images were labelled as normal. Indicatively, Figure 1 presents four such images.

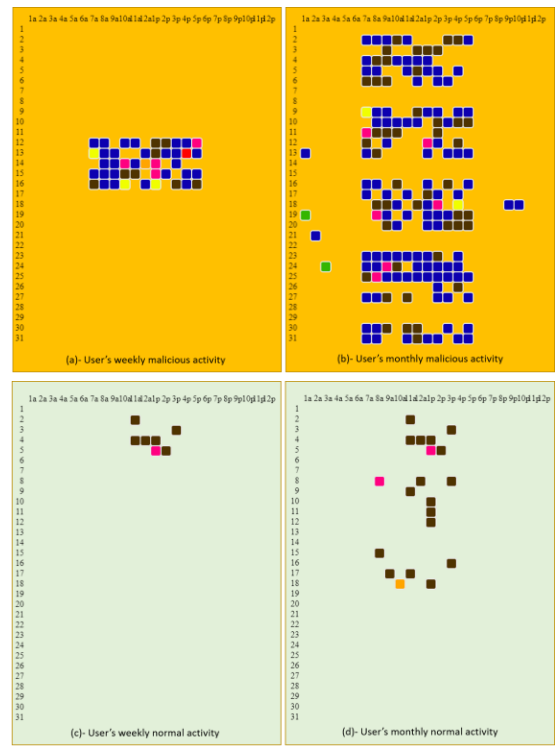

Fig. 1. Display of user' normal and malicious activity weekly and monthly

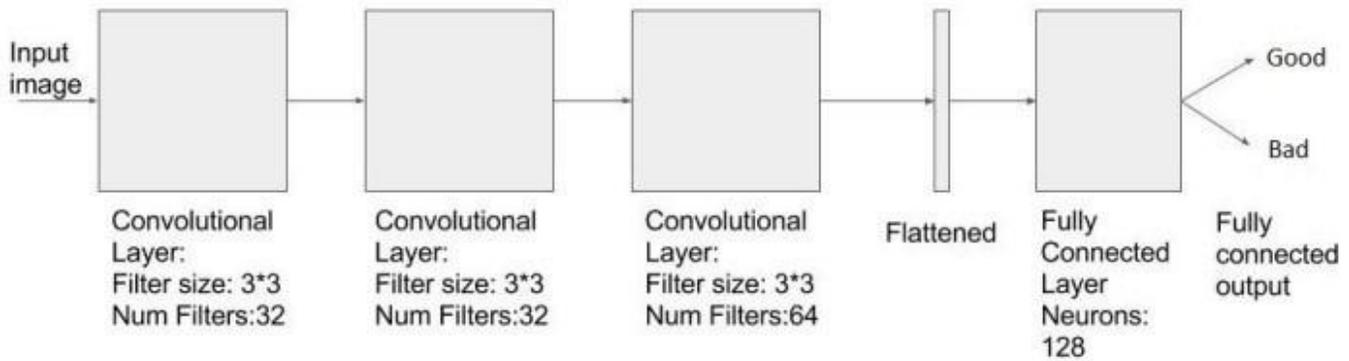

Fig. 2. Schematic layout of the CNN algorithm implemented

*3) Third Stage:* In the third stage of the experiment, Google's Tensorflow program was used to design a six-layer CNN that would recognize whether the image was classified as normal or malicious. CNN is a class of forward-facing artificial neural networks and has been successfully applied to the analysis and recognition of visual images, videos, and natural language processing systems. It also uses relatively little processing compared to other image sorting algorithms. This means that the network easily learns filters made using traditional algorithms. It is also known as an invariant artificial neural network, based on the weight's architecture. Finally, in the previous step one thousand one hundred ninety-nine (1199) images were produced, of which - as reported - four hundred and thirty (430) were assessed as containing normal activity. For this reason, a corresponding number of malicious images were selected. Finally, eight hundred and sixty (860) images were used for this stage, of which eight hundred forty (840) were the training data images as follows:

1) *Training Data*: 80% of training images were used.
2) *Validation Data*: 20% of training images were used for validation.

Figure 2 shows the resulting implementation.

*B. Evaluation*

In order to ensure a balanced dataset, the 769 malicious activity images were reduced through random selection to 430 images, matching the number of normal activity images. The resulting dataset had 860 samples, including an equal number of normal and malicious activity samples; the dataset was split as follows: 20 images were set aside for validation and the remaining 840 images were split 80% training (672 images) and 20% testing (168 images). The validation images were also selected in a balanced fashion, with an equal number (10) of normal and malicious activity samples. The forecasting of the results was successful, with a proportion that reached 100%; the CNN algorithm was traced graphically to illustrate training accuracy, validation accuracy, and cost. Below are presented the three training exercises (training accuracy, validation accuracy, cost) and the implementation graph of the CNN algorithm.

According to the graphs in Figure 3, training accuracy starts at a value close to 0.450, i.e., 45%, and has an increasing value to reach a rate close to 100%. Validation accuracy (Figure 3) starts from a value close to 0.400, i.e., 40%, and has a rising value to reach a rate close to 90%. While the cost starts froma value close to 0.700, i.e., 70% and has a decreasing value to arrive at a value, which is close to 0.

## IV. CONCLUSION

The purpose of this study was to investigate the feasibility of using machine learning techniques to detect malicious activity by converting activity reports into a visual representation. The answer to this question has gone through three stages: a) the collection, processing, and classification of the data of the users under consideration, b) the visualization of the extracted data, and c) the use of the CNN algorithm to classify behaviour into malicious or normal. The algorithm was trained and tested using a dataset of 860 created images, including both malicious and normal activity. Our conclusion is that, with the methodology used, the malicious activity of the users of the information system was achieved. The forecasting of the results was successful, with a percentage that reached 100%.

Table I is a concise table showing the results of some internal recognition methods, both in terms of validation accuracy and some of their comparisons.

In conclusion, it should be noted that the characterization of a user's behaviour as malicious is also dependent on the Security Policy that the Company or Organization adopts to protect its Information System. For the analysis, one of the sets polices was that visiting social networking sites or job search websites is falls outside normal activity and should be categorised as malicious. In addition, an important role in characterising a user's activity is played by the position held in the organization and the set policy and associated analysis should also consider this.

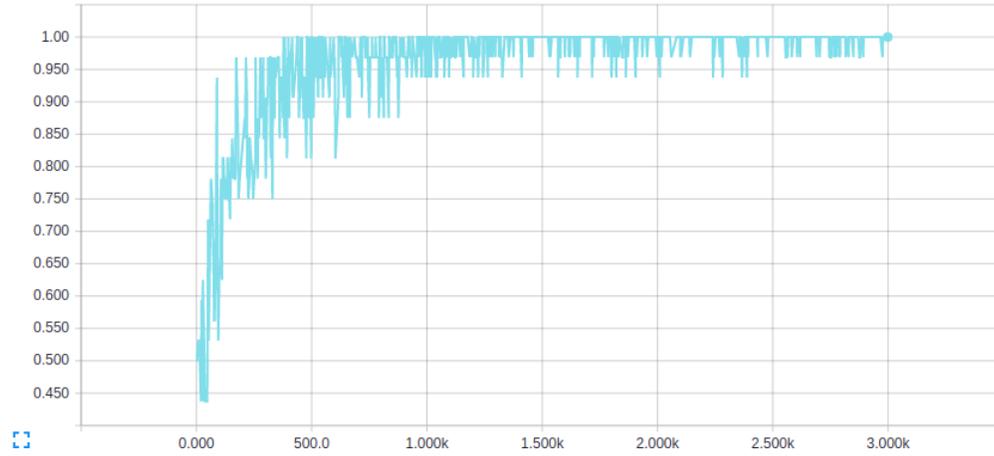

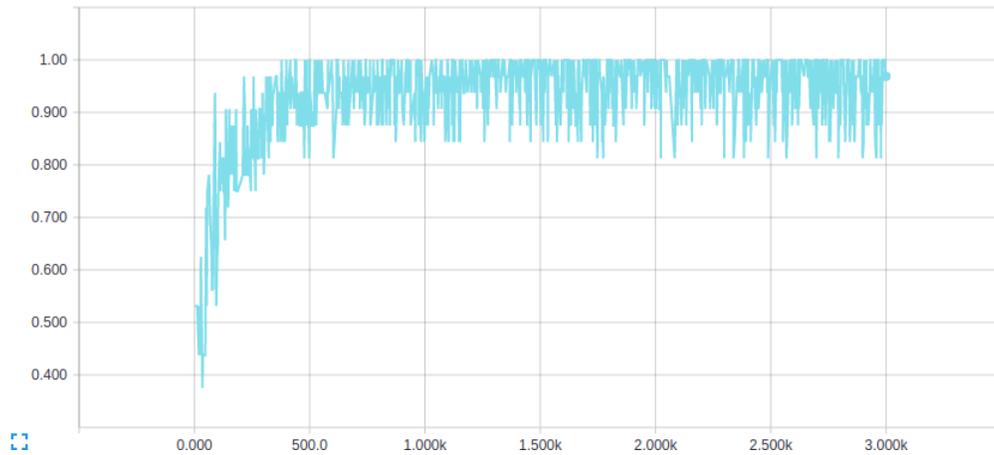

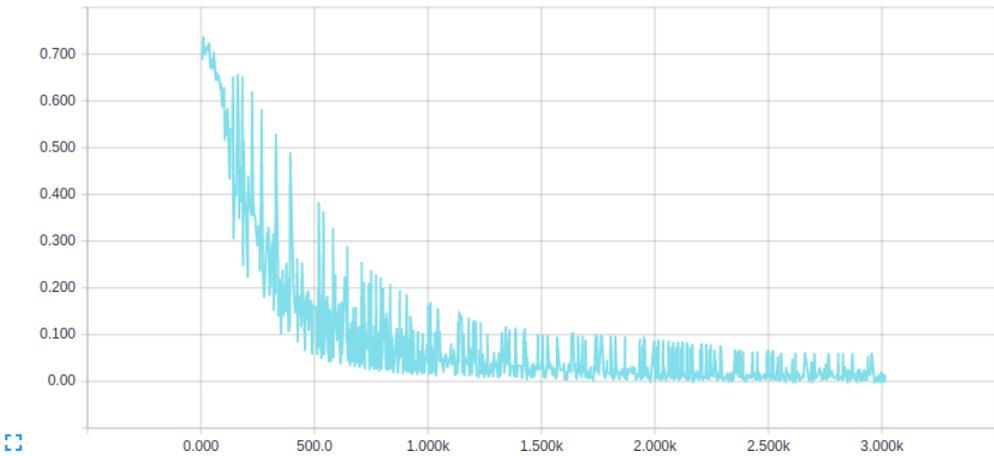

Fig. 3. Graphic representation of Training Accuracy, Validation Accuracy and Cost

TABLE I
COMPARISON OF OUR METHOD WITH OTHERS

| Study | Accuracy | Comparison |
|---|---|---|
| Proposed method | Testing Accuracy: 100% Training Accuracy:100% Validation Accuracy:90.6% Cost: 0.582 | Mechanical learning (machine learning) and training of the CNN algorithm for the implementation and categorization of the user's behaviour in normal and malicious. |
| W. Eberle et al [11] | Testing Accuracy: 95% | In this method a graphical theoretical approach was used to detect malicious user behaviour in an Information System |
| O. Brdiczka et al [12] | Server Eitrigg:82.74% Server Cenarion Circle:89.09% Server Bleeding Hollow:79.84% | This study proposes an approach combining structural anomaly detection (SA) from social networks and information networks as well as the psychological profiling (PP) of individuals. |
| W. T. Young et al [13] | Indicators:87.4% Anomalies:97.9% Scenarios:80.6% | This study focuses on the awareness of domain knowledge to detect malicious behaviour through the use of specific SAs algorithms. |
| J. Breier et al [14] | | The error rate of the method used was less than 10%. In the method we used, the error rate approaches zero (0). |
| P. A. Legg et al [10] | | The method used was based on collecting log data, by building a profile for each user and his / her property. In the method we used, the user property was not evaluated. |
| X.Ren, L.Wang [3] | | The method proposes a hybrid intelligent system for insider threat detection that aims to realize more effective detection of security incidents by incorporating multiple complementary detection techniques, such as entity portrait, rule matching and iterative attention. In our method, only the user activity was evaluated |
| A.Sajjanhar et al [4] | Accuracy 99% | The method proposes an image-based feature representation of the daily resource usage pattern of users in the organization. Classification models are applied to the representative images to detect anomalous behaviour of insiders. The images are classified too malicious and no malicious |
| J.Kim et al [5] | Accuracy > 90% | The method proposes insider-threat detection methods based on user behaviour modelling and anomaly detection algorithms. Experimental results show that the proposed framework can work reasonably well to detect insiders' malicious behaviours. Although the proposed framework was empirically verified, there are some limitations in the current research, which led the authors to future research directions |

ACKNOWLEDGMENT

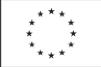

This project has received funding from the European Union's Horizon 2020 research and innovation programme under grant agreement no. 786698. The work reflects only the authors' view and the Agency is not responsible for any use that may be made of the information it contains.